
\input amstex
\magnification=1200
\TagsOnRight
\def\wh{\widehat}
\def\wt{\widetilde}

\def\ov{\overline}
\def\np{\not{\text{P}}}

\def\ep{\varepsilon}
\nopagenumbers
\headline={\ifnum\pageno=1\hfil\else\hss\tenrm -- \folio\ --\hss\fi}
\line{Preprint {\bf SB/F/94-222}}
\hrule
\vskip 2,5cm
\font\nine=cmr9
\centerline{\bf{CANONICAL COVARIANT FORMULATION OF GREEN-SCHWARZ}}
\centerline{\bf {SUPERSTRING WITHOUT SECOND CLASS CONSTRAINTS}}
\vskip 2cm
\centerline{\it{A. Restuccia}}
\vskip 3mm
\centerline{\it{and}}
\vskip 3mm
\centerline{\it{J. Stephany}}
\vskip 3mm
\centerline{\it{Universidad Sim\'on Bol\'{\i}var}}
\centerline{\it{Departamento de F\'{\i}sica}}
\centerline{\it{Apartado Postal 89000, Caracas 1080-A}}
\centerline{\it e-mail: stephany{\@}usb.ve}
\vskip 1cm
{\nine
\centerline{\bf Abstract}
\vskip 3mm
{\narrower{\flushpar
We describe a canonical covariant formulation of the
Green-Schwarz Superstring which allows the construction of a new
covariant action canonically equivalent to the Green-Schwarz action but
subjected only to first class constraints. From this action the correct  BRST
operator for the quantization of the Green-Schwarz Superstring may be
constructed. Also the gauge fixed action   in the  Light-Cone
gauge may be reobtained. The action presented in this letter generalizes in a
non-trivial form the action introduced by Kallosh for the
Brink-Schwarz-Casalbuoni Superparticle.\par}}} \vskip 3cm

\hrule
\bigskip
\centerline{\bf UNIVERSIDAD SIMON BOLIVAR}

\newpage

The advance in the development of Superstring theory has
been delayed for some time by  the lack of an explicitly
covariant quantization environment in which the second quantized theory of
interacting Superstrings could be discussed . For the
Green-Schwarz Superstring (GSS)[1] this problem reveals itself through
the unsuccessful result of the various attempts to
construct the correct BRST operator of the theory. The main
difficulty is related to the fact that the first class
constraints associated to the local $\kappa$-supersymmetry
[2] appear mixed with second class constraints in such a way
that no  local and Lorentz covariant quantization of the
system appears to be allowed[3]. On the other hand the definite success of
the light cone gauge approach for the computation of finite multiloop
amplitudes [4] reaffirm our expectations of having a theory of
fundamental interactions in terms of  Superstrings.
Moreover recent results for the bosonic case [5] renew our
confidence in the feasibility of  a covariant second quantized
Superstring theory once we have constructed the correct BRST operator in the
first quantized formulation.

The zero mode structure of GSS is described by the Brink-Schwarz and
Casalbuoni (BSCS) superparticle [6] which in particular has a constraint
structure which is similar to, although simpler, than the one of the GSS.
After many attemps the correct BRST operator for the BSCS was finally
constructed in  Ref.[7]  in terms of
an infinite set of auxiliary fields. Later in Ref.[8] Kallosh was able to
construct an action from which the previously constructed BRST operator may be
deduced. This resolves the problem for the BSCS but regretfully this
construction has no obvious generalization for the GSS. Another approach to
the BSCS using twistor variables may be found in Ref.[9] (see also references
in Ref.[10]).   In a recent paper[10] we presented a canonical covariant
approach which, starting from the original BSCS action
 and  by enlarging the phase space with the introduction of appropriate
auxiliary variables [11], allowed the construction of the  action presented by
Kallosh in a systematic way. In this paper we present the generalization
of this approach to the GSS obtaining a local, Lorentz covariant action
which generalizes Kallosh action in a non-trivial way. From this action
the correct BRST operator needed for the quantization of the GSS could
be obtained using standard methods.      .

The Green-Schwarz  action, for type $IIB$ Superstring, [1] is given
$$
S(x,\theta )=<L_1+L_2>\tag 1a
$$
where
$$\align
L_1 &=-\frac{1}{2}\sqrt{-g}g^{\alpha\beta}\pi^\mu_\alpha\pi_{\mu\beta}\\
L_2 &=-\ep^{\alpha\beta}\partial_\alpha x^\mu (k^1_{\mu\beta}-
k^2_{\mu\beta})-\ep^{\alpha\beta}k^{1\mu}{}_{\alpha}k^2{}_{\mu\beta}\\
\pi_\mu^\alpha &=\partial_\alpha x^\mu-i\ov{\theta}^A\gamma^\mu\partial_\alpha
\theta^A\\
k^{A\mu}{}_\alpha &=i\ov{\theta}^A\gamma^\mu\partial_\alpha\theta^A \ \ {\text
{without sum}}\ \ \,\ \  A=1,2 . \tag 1b
\endalign
$$
Here $\theta^A=1,2$, are 10-dim Majorana-Weyl spinors of the same chirality.
We denote with $<>$ integration on the 2-dim world sheet variables
$\tau$ and $\sigma$. The world sheet indices are $\alpha$ and $\beta$  while
$\mu$ and $\nu$ denote the 10-dim  target space indices.
Introducing $\eta^A$ and $P_\mu$   the conjugate momenta to $\theta^A$ and
$x^\mu$
the canonical analysis [3] yields the  constraints (using primes to denote
$\partial_{\sigma}$)
$$\align
\varphi_- &\equiv \frac{1}{2}(P_\mu -x_\mu{}')^2-2\eta^1\theta^1{}',\\
\varphi_+ &\equiv \frac{1}{2}(P_\mu -x_\mu{}')^2+2\eta^2\theta^2{}', \tag 2a
\endalign
$$
which are first class and the constraints
$$
\align
F_- &\equiv \eta^1+i\ov{\theta}^1\gamma^\mu (P_\mu -x_\mu{}'+
k^1_{\mu\sigma})=0  , \\
F_+ &\equiv \eta^2+i\ov{\theta}^2\gamma^\mu (P_\mu -x_\mu{}
'+k^2_{\mu\sigma})=0  ,
 \tag 2b
\endalign
$$
which are a mixture of first and second class ones.  The latter (2b) may be
covariantly decomposed into first class contraints
$$
\psi_- \equiv F_-\Gamma_-=0\ \ \ ,\ \ \ \psi_+ \equiv F_+\Gamma_+=0,\tag 3a
$$
and second class ones
$$
F_-\Gamma_+=0 \ \ \ ,\ \ \ F_+\Gamma_-=0, \tag 3b
$$
by using the following definitions and properties of the $\Gamma_+$,
$\Gamma_-$ matrices
$$\align
\Gamma_- &\equiv \gamma^\mu (P_\mu-x_\mu '+2k^1_{\mu\sigma})\\
\Gamma_+ &\equiv \gamma^\mu (P_\mu-x_\mu{}'-2k^2_{\mu\sigma}), \tag 4\\
\Gamma_-\Gamma_- &\equiv 2H_-{\text{\bf 1}}
=2(\varphi_-+2F_-\theta^{1'}){\text{\bf 1}}\\
\Gamma_+\Gamma_+ &\equiv 2H_+{\text{\bf 1}}
=2(\varphi_+-2F_+\theta^{2'}){\text{\bf 1}}\\
\Gamma_-\Gamma_+=\Gamma_+\Gamma_-&=2(P^2-x^{'2}-2k^2_{\mu\sigma}
(P^\mu-x^{'\mu})+\\
&+2k^1_{\mu\sigma}(P^\mu +x^{'\mu})-4k^1_{\mu\sigma}k^{2\mu}_{}\sigma
){\text{\bf 1}}.\tag 5
\endalign
$$
Nevertheless $\psi_+$ and $\psi_-$ are infinite reducible
constraints and this together with the fact that (3b) are second class are
the obstacles mentioned above to the quantization of the system.

We observe that the $-(+)$ constraints in (2) are associated to the left
(right) moving  sector of the Superstring which decouple. In particular
$\Gamma_-$, $F_-$, $\varphi_-$, $F_-\Gamma_-$ commute, under Poisson bracket,
with $\Gamma_+$, $F_+$, $\varphi_+$, $F_+\Gamma_+$.

Following our strategy in Ref. 10 we are now going to  extend the phase space
in order to eliminate the second class constraints.  To this end we will
construct in the extended phase space a new dynamical  system restricted only
by first class constraints such that with an admissible  partial gauge fixing
it reduces to the original system, with the correct  quantum measure [12]. The
procedure starts, as in the case of the superparticle,  by constructing a new
dynamical system $S_n$, with $n$-stages of reducibility,  canonically
equivalent to the GSS. Then we consider the limit $n\to \infty$ in a sense
that  will be precisely defined and show that in this limit the  new
dynamical system contains only first class constraints.

We are going to treat the left sector explicitly. The procedure for the right
sector is similar. We introduce the new canonical variables $\eta_1$ and
$\xi_1$ and consider the  extension of $F_-$ given by
$$
\wt{F}_-\equiv F_-+\Phi_1  . \tag 6
$$
with
$$\align
\Phi_1 &=\eta_1+\ov{\xi}\cdot \omega_1,\tag 7\\
\ov{\xi}\cdot \omega_1 &\equiv <\ov{\xi}_1(\wh{\sigma})
\omega_1(\wh{\sigma},\sigma )>_{\wh{\sigma}} .\tag 8
\endalign
$$
Here $<>_{\wh{\sigma}}$ denotes integration on $\wh{\sigma}$ and the matrix
$\omega_1=\omega_1(\wh{\sigma},\sigma )$ is independent of $\eta$ and $\eta_1$.

We determine $\omega_1$ from the condition
$$
\{\wt{F}_-(\sigma ),\wt{F}_-(\wh{\sigma})\}=0 . \tag 9
$$
The solution to this problem is given by
$$\align
\omega_1(\sigma ,\wh{\sigma}) &=-i\Gamma_0\delta (\sigma ,\wh{\sigma})-
\frac{4}{3}(\ov{\xi}_1\gamma^\mu \xi_0{}')\gamma^\mu
\delta (\sigma ,\wh{\sigma})+\\
&+\frac{1}{3}(\ov{\xi}_1(\sigma )\gamma^\mu \xi_1(\wh{\sigma})\gamma_\mu
\frac{\partial}{\partial\sigma}\delta (\sigma ,\wh{\sigma}),\tag 10
\endalign
$$
where $\Gamma_0\equiv \Gamma_-$ and $\xi_0\equiv \theta^1$.

It is a non trivial generalization of the solution $\omega =-\np$ for the
similar problem in the case of the superparticle [10]. It depends not only  on
the original canonical variables but also in the new one $\xi_1$.

We now introduce the new geometrical object
$$
\wh{\Phi}_1\equiv \eta_1+\ov{\xi}_1\cdot W_1 ,\tag 11
$$
with $W_1(\wh{\sigma},\sigma )$ is determined from the condition
$$
\{\wt{F}_-(\sigma ),\wh{\Phi}(\wh{\sigma})\}=0 . \tag 12
$$
There exists a solution to this equation given by
$$\align
W_1(\sigma ,\wh{\sigma}) &=i\Gamma_0\delta (\sigma ,\wh{\sigma})+\frac{8}{3}
(\ov{\xi}_1\gamma^\mu \xi_0)\gamma^\mu \delta (\sigma ,\wh{\sigma})+\\
&-(\ov{\xi}_1(\sigma )\gamma^\mu \xi_1(\wh{\sigma})\gamma_\mu
\frac{\partial}{\partial\sigma}\delta (\sigma ,\wh{\sigma}),\tag 13
\endalign
$$
which again is a generalization of the $W_1=\np$ introduced for the analysis
of the superparticle [10].

We now obtain the appropriate extension of the projectors (4). We define
$\Gamma_1(\sigma )$ through
$$
\{\wh{\Phi}(\sigma ),\wh{\Phi}(\wh{\sigma})\}= 2i\Gamma_1(\sigma )
\delta (\sigma ,\wh{\sigma}) . \tag 14
$$
$\Gamma_1(\sigma )$ is then given by
$$
\Gamma_1(\sigma )=\Gamma_0(\sigma )-4i(\ov{\xi}_1\gamma_\mu \xi_0{}')
\gamma^\mu+2i(\ov{\xi}_1\gamma_\mu \xi_1{}')\gamma^\mu \tag 15
$$
satisfying
$$\align
\Gamma_1\Gamma_1 &=2H_1 {\text{\bf 1}},\\
H_1 &\equiv \frac{1}{2}(P_\mu-x'_\mu+2k^1_{\mu\sigma}-4i
(\ov{\xi}_1\gamma_\mu \xi_0{}')+2i(\ov{\xi}_1\gamma_\mu \xi_1{}'))^2 . \tag16
\endalign
$$
The extension of $\varphi_-$ is then defined by
$$
\varphi_1=H_1-2\wh{\Phi}_1(\xi_1'-\xi_0') . \tag 17
$$
It satisfies the following commutation relations
$$\align
\{\varphi_1(\sigma ),\varphi_1(\wh{\sigma})\} & = 2
(\varphi_1(\sigma )
+ \varphi_1(\wh{\sigma}))\frac{\partial}{\partial\wh{\sigma}} \delta (\sigma
,\wh{\sigma})\\ \{\varphi_1(\sigma ),\wh{\Phi}(\wh{\sigma}) & = -2
\wh{\Phi}(\sigma )\frac{\partial}{\partial\wh{\sigma}}
\delta (\sigma ,\wh{\sigma})\\
\{\varphi_1(\sigma ),\wt{F}_-(\wh{\sigma})\} &=0 .\tag 18
\endalign
$$
The original set of constraints (2) is now reformulated in the extended
phase space in the following way
$$\align
\wt{F}_- &=F_-+\Phi_1=0 \tag 19a\\
\varphi_1 &=0 \tag 19b\\
\wh{\Phi}_1 &=0 . \tag 20
\endalign
$$
 Constraints (19)  are first class,  while (20) is still a
mixture of first  and second class constraints. The equations above correspond
to the left moving sector, there is an  analogous set for the right moving
sector.

The first class part of $\wh{\Phi}_1$ may be decoupled by considering
$$
\align
\psi_1 &\equiv (F_-+\Phi_1-\wh{\Phi}_1)\Gamma_1=0 , \tag 21a\\
\wh{\Phi}_1^T &\equiv \wh{\Phi}_1\Gamma_+=0. \tag 21b
\endalign
$$
which is are the generalizations of equations (3).
We have then the constraints
$$\align
\wt{F}_-&=0, \tag 22a\\
\varphi_1 &=0, \tag 22b\\
\psi_1 &=0, \tag 22c\\
\wh{\Phi}_1^T &=0. \tag 23
\endalign
$$
Again constraints (22) are first class while constraints (23) are second
class. The constraints $\psi_1$ are infinite reducible.  We may recover the
original set of left moving constraints in (2) by doing a partial  gauge
fixing. The reduction is performed by considering the partial  gauge fixing
condition, associated to (22a),  $$\align  \wt{\chi}_-^L &\equiv
\Gamma_+\xi_1=0 \tag 24a\\ \wt{\chi}_-^T &\equiv \eta_1\Gamma_+=0 . \tag 24b
\endalign
$$
{}From (24) and (22) we obtain $\eta_1=0$, $\xi_1=0$. Eq.(22) then reduces
exactly  to the left moving constraints in (2), (3). Moreover the reduction
may also  be obtained in the functional integral, with the correct quantum
measure [12].  The proof goes in the same way as for the superparticle
extension in [10].

Having constructed $S_1$ we may now proceed to obtain $S_n$. We introduce
$$
\Phi_i\equiv \eta_i+\ov{\xi}_i\cdot \omega_i ,\ \ \ \ \ \ \ \ \ i=1,\cdots ,n
\tag 25
$$
where
$$\align
\omega_1(\sigma ,\wh{\sigma}) &=-i\Gamma_{i-1}\delta (\sigma ,\wh{\sigma})-
\frac{4}{3}(\ov{\xi}_i\gamma^\mu (\xi{}'_{i-1}-\xi{}'_{i-2}+\cdots
\xi{}'_0))\gamma^\mu \delta (\sigma ,\wh{\sigma})+\\
&+\frac{1}{3}(\ov{\xi}_i(\sigma )\gamma^\mu \xi_i(\wh{\sigma})\gamma_\mu
\frac{\partial}{\partial\sigma}\delta (\sigma ,\wh{\sigma}),\tag 26
\endalign
$$
and
$$
\wh{\Phi}_i \equiv \eta_i+\ov{\xi}_i\cdot W_i \tag 27
$$
with
$$\align
W_i(\sigma ,\wh{\sigma}) &=i\Gamma_{i-1}\delta (\sigma ,\wh{\sigma})+
\frac{8}{3}(\ov{\xi}_i\gamma^\mu (\xi{}'_{i-1}-\xi{}'_{i-2}+\cdots \xi{}'_0))
\gamma^\mu \delta (\sigma ,\wh{\sigma})+\\
&-(\ov{\xi}_i(\sigma )\gamma^\mu \xi_i(\wh{\sigma})\gamma_\mu
\frac{\partial}{\partial\sigma}\delta (\sigma ,\wh{\sigma}) . \tag 28
\endalign
$$
We now introduce $\Gamma_i(\sigma )$ generalizing (14) through the equation
$$
\{\wh{\Phi}_i(\sigma ),\wh{\Phi}_i(\wh{\sigma})\}=2i \Gamma_i(\sigma )
\delta (\sigma ,\wh{\sigma}), \tag 29
$$
which give
$$
\Gamma_i(\sigma )=\Gamma_{i-1}(\sigma )-4i(\ov{\xi}_1\gamma_\mu
(\xi{}'_{i-1} - \xi{}'_{i-2} + \cdots \xi{}'_0))
\gamma^\mu+2i(\ov{\xi}_i\gamma_\mu \xi{}'_i)\gamma^\mu  . \tag 30
$$
The expressions for $H_n$ and $\varphi_n$ are obtained from
$$\align
\Gamma_n\Gamma_n &=2H_n {\text{\bf 1}},\\
H_n &\equiv \frac{1}{2}(P_\mu -x'_\mu+2k^1_{\mu\sigma}-4i
(\ov{\xi}_n\gamma_\mu (\xi{}'_{n-1}-\xi'_{n-2}+\cdots \xi{}'_0))+\\
&+2i(\ov{\xi}_n\gamma_\mu \xi{}'_n))^2\\
\varphi_n &=H_n-2\wh{\Phi}_n(\xi{}'_n-\xi{}'_{n-1}+\cdots \xi{}'_0). \tag 31
\endalign
$$
Finally we consider
$$\align
\psi_n &\equiv \Psi_n\Gamma_n, \tag 32a \\
\Psi_n &\equiv F_-+\Phi_1-\wh{\Phi}_1-\Phi_2-\wh{\Phi}_2+\Phi_3-\cdots
\wh{\Phi}_n=\\
 &=F_-+\ov{\xi}_1(\omega_1-W_1)-\ov{\xi}_2(\omega_2-W_2)+\cdots
\ov{\xi}_n(\omega_n-W_n).\tag 32a
\endalign
$$
The complete set of constraints associated to $S_n$ are
$$\align
F_-+\Phi_1 &=0\\
\wh{\Phi}_1+\Phi_2 &=0\\
\cdots \cdots \cdots & \cdots \cdots \\
\wh{\Phi}_{n-1}+\Phi_n &=0 \tag 33a\\
\varphi_n &=0 \tag 33b\\
\psi_n &=0 \tag 33c\\
{\Psi}_n^T &=0 \tag 34
\endalign
$$
The constraints (33) are  first class with $\psi_n$  infinite
reducible while
$$
{\Psi}_n^T \equiv {\Psi}_n\Gamma_+ \tag 35
$$
are second class constraints. We note that the constraint (34) is equivalent
to  ${\wh\Phi_n}^T=0$.  We may again recover the
GSS in the original phase space restricted by the set of constraints in (2) by
doing a partial gauge  fixing. The reduction is performed by considering the
following gauge  fixing conditions, associated to (33a), $$\align
\Gamma_+\xi_1=0\ \ \ &,\ \ \ \eta_1 \Gamma_+=0\\ \Gamma_+\xi_2=0\ \ \ &,\ \ \
\ \Gamma_-\xi_1=0\\ \Gamma_+\xi_3=0\ \ \ &,\ \ \ \ \Gamma_-\xi_2=0\\
\cdots \cdots \cdots \cdots &,\cdots \cdots \cdots \cdots \cdots \\
\Gamma_+\xi_n=0\ \ \ &,\ \ \ \ \Gamma_-\xi_{n-1}=0 . \tag 36
\endalign
$$
{}From (36) we obtain
$$\align
\xi_1 &=0,\\
\xi_2 &=0,\\
\cdots \cdots &\cdots \cdots ,\\
\xi_{n-1} &=0,\tag 37a\\
\Gamma_+\xi_n &=0 .\tag 37b
\endalign
$$
{}From the transverse projection of (33a),  multiplying then by
$\Gamma_+$, and using (34) we get
$$
\Gamma_-\xi_n =0 \tag 38
$$
which together with (37b) yield
$$
\xi_n =0 . \tag 39
$$
{}From (33a) and (33c) we then obtain
$$\align
\eta_1 &=0\\
\cdots &\cdots \cdots \\
\eta_n &=0 \tag 40
\endalign
$$
The system defined by $S_n$, restricted  by (33) and (34)  then
reduces to original system subjected to the  constraints in (2). The reduction
 in the functional integral, is  performed with the correct quantum measure
[11][12]. The generalization needed to include the right moving sector
presents no further complications.

 The constrained system
$S_\infty$ is determined from (33) taking  $n\to \infty$. The key point is
that in this limit there are only first class constraints.
This can be shown by taking the gauge fixing conditions
$$
\Gamma_+\xi_{i+1}=0 \ \ \ ,\ \ \ \Gamma_-\xi_i =0 \tag 41
$$
which generalize (36).For any $i\geq n$, system $S_\infty$ reduces to $S_n$
and hence is equivalent  to the original constrained system. Moreover one can
show that in this gauge $\Psi^T_n$ is satisfied identically and that the
algebra of the first class constraints (33) closes without further
restrictions.
The action associated to $S_\infty$ may then be written in the form
$$\align
S_\infty &<P_\mu \dot{x}^\mu +S^-_\infty +S^+_\infty >,\tag 42a\\
S^-_\infty &= <\sum^\infty_{i=0}\eta_i\dot{\xi}_i+\alpha \varphi_\infty +
\beta\psi_\infty +\\
&\sum^\infty_{i=0}\lambda_i(\wh{\Phi}_i+\Phi_{i+1})>\tag 42b
\endalign
$$
where $\eta_0\equiv \eta$, $\xi_0\equiv \theta^1$, $\wh{\Phi}_0\equiv F_-$ and
$S^+_\infty$ is the analogous construction for the right moving sector, in
terms of new independent auxiliary fields. The system  (42) is infinite
reducible as a consequence of the infinite reducibility of $\psi_\infty$.
The others constraints are irreducible. $S^+_\infty$ is obtained from the
expression of $S^-_\infty$ changing the sign of the terms with a
$\partial_\sigma$ derivative.

We have thus constructed a new action given by (42) constrained only by first
class constraints which has the same degrees of freedom and is canonically
equivalent to the original Green-Schwarz action. This formulation is in terms
of regular first class constraints only allowing a consistent construction of
the BRST charge with the right cohomology for Superstring.The
systems $S_n$ in our approach are  still  restricted by some second
class constraints which are needed in order to match the degrees of
freedom of the  Green-Schwarz Superstring. In the
limit $n \to \infty$  only first class constraints appear but we still obtain
the correct number of degrees of freedom. As in the superparticle case [10]
the manipulation of the infinite auxiliary fields as well as the infinite
tower of ghosts for ghosts which appear due to the infinite reducibility of
the system, may require the introduction of generating functions [14]. In this
line of thought it is interesting to notice that  gauge fixing conditions
which are equivalent  for each one of the $S_n$ actions may
 be inequivalent in the  $n \to \infty$ limit.

\vskip .3cm
\item{}{\bf References}
\vskip .3cm

\item{[1]} M. Green and J. Schwarz, {\it Phys Lett} {\bf 136B} (1984) 367;
{\it Nucl Phys}{\bf 243} (1984) 475.
\item{[2]} W. Siegel, {\it Phys Lett} {\bf 128B} (1983) 397.
\item{[3]} T. Hori and K. Kamimura, {\it Prog Theor Phys} {\bf 73} (1985) 476.
\item{[4]} A. Restuccia and J. G. Taylor, {\it Int J of Mod Phys} {\bf A8}
(1993) 753; {\it Phys Lett} {\bf 282B} (1992) 377; {\it
Phys. Reports} {\bf 174} (1989) 285; Preprint KCL (1993); P.diVechia et al
{\it Nucl Phys} {\bf B298} (1988) 527; E.D+Hoker and D.H.Phong {\it Comm Math
Phys} {\bf 125} (1989) 469; S.Mandelstam {\it Phys Lett} {\bf B277} (1992) 82;
N.Berkovits {\it Nucl Phys} {\bf B395} (1993) 77;  {\it Phys Lett} {\bf B300}
(1993) 53.     \item{[5]} B. Zwiebach {\it Nucl Phys} {\bf B390} (1993) 33.
\item{[6]} R.Casalbuoni {\it Nuov Cim} {\bf 33A} (1976) 389; L. Brink and J.
Schwarz, {\it Phys Lett} {\bf 100B} (1981) 310.
\item{[7]} E. Bergshoeff and R.Kallosh, {\it Phys Lett} {\bf B240} (1990)
105.
\item{[8]} R. Kallosh, {\it Phys Lett} {\bf B251} (1990) 134.
\item{[9]}Y Eisenberg and S.Solomon, {\it Nuc Phys}  {\bf B309} (1988) 709;
N.Berkovits, {\it Nuc Phys}  {\bf B350} (1991) 193; Y.Eisenberg, Preprint
IASSNS-HEP-91/48, Princeton; E.Nissimov, S.Pacheva and S.Solomon {\it Nucl
Phys}{\bf B317} (1989) 344.
\item{[10]} A. Restuccia and J.Stephany, {\it Phys
Rev } {\bf D47} (1993) 3437.
\item{[11]} A. Restuccia and J.Stephany, {\it Phys Lett} {\bf 305B} (1993) 348;
R. Gianvittorio, A. Restuccia and J.Stephany, {\it Mod Phys  Lett } {\bf A6}
(1991) 2121; I.Batalin and E.S.Fradkin, {\it Nucl Phys} {\bf 279} (1987) 514.
131.
\item{[12]} G. Senjanovic, {\it Ann Phys} {\bf 100} (1976) 227.
\item {[13]} A.Restuccia and J.Stephany in preparation
.\item{[14]} A.Galperin et al,  {\it
Class Quantum Grav} {\bf 1} (1984) 469;  I.Bars and R.Kallosh, {\it Phys Lett}
{\bf B233} (1989) 117; M.B.Green and C.M.Hull, {\it Phys Lett} {\bf B229}
(1989) 215.
\bye